\documentclass[a4paper,11pt]{article}
\pdfoutput=1 

\usepackage{jcappub} 
\usepackage{url}
\usepackage[T1]{fontenc} 
\usepackage{amsmath}
\usepackage{booktabs} 
\usepackage{threeparttablex}
\usepackage{graphicx}
\usepackage{dcolumn}
\usepackage{bm}
\usepackage{amssymb}
\usepackage{times}
\usepackage{threeparttable}
\usepackage{orcidlink}

\title{\boldmath Seyfert Galaxies as Neutrino Sources: An Outflow–Cloud Interaction Perspective}

\author[a]{Zhi-Peng Ma\,\orcidlink{0009-0007-0717-3667}}
\author[a,1]{Kai Wang\,\orcidlink{0000-0003-4976-4098}}
\author[a]{Yuan-Yuan Zuo\,\orcidlink{0009-0005-5745-9466}}
\author[a]{Yuan-Chuan Zou\,\orcidlink{0000-0002-5400-3261}}

\affiliation[a]{Department of Astronomy, School of Physics, Huazhong University of Science and Technology, Wuhan, Hubei 430074, China}

\emailAdd{kaiwang@hust.edu.cn}

\abstract{Following the identification of the first confirmed individual neutrino source, Seyfert galaxies have emerged as the most prominent class of high-energy neutrino emitters. In this work, we perform a detailed investigation of the outflow--cloud interaction scenario for neutrino production in Seyfert nuclei. In this framework, fast AGN-driven winds collide with clumpy gas clouds in the nuclear region, forming bow shocks that efficiently accelerate cosmic-ray protons. The accelerated protons subsequently interact with cold protons from the outflows via inelastic proton--proton (\(pp\)) collisions, producing high-energy neutrinos, while the photomeson (\(p\gamma\)) process with disk photons may provide a subdominant contribution at the highest energies. Applying this model to five neutrino-associated Seyfert galaxies, we successfully reproduce the observed TeV neutrino fluxes without violating existing gamma-ray constraints. By integrating over the Seyfert population using X-ray luminosity functions, we further demonstrate that Seyfert galaxies can account for a substantial fraction of the diffuse astrophysical neutrino background in the \(10^{4}\)--\(10^{5}\,\mathrm{GeV}\) energy range.}

\begin{document}
\maketitle
\flushbottom

\section{Introduction}

High-energy neutrinos provide a unique probe of nonthermal processes in the universe. Since the first detection of a diffuse neutrino flux in the 10--$10^3$~TeV range by the IceCube Collaboration in 2013~\cite{icecube2013evidence}, the sources responsible for these fluxes have remained largely unidentified. In 2022, however, the IceCube Collaboration reported compelling evidence for the emission of 1--10~TeV neutrinos from the Seyfert~II galaxy NGC~1068 at a significance level of $4.2\sigma$~\cite{icecube2022evidence}. A subsequent analysis further increased the local significance to $5.0\sigma$, establishing NGC~1068 as a firmly confirmed neutrino source~\cite{abbasi2025icecube}. In addition to NGC~1068, the recently identified NGC~7469 
\noindent shows the second strongest signal, with a local significance of $3.8\sigma$~\cite{abbasi2025icecube}; this galaxy had previously been proposed to coincide spatially with two $\sim100$~TeV neutrino events~\cite{sommani2025two}. Furthermore, follow-up IceCube searches for neutrino emission from X-ray–bright AGN have revealed suggestive excesses (at $\lesssim3\sigma$ significance) from several other Seyfert galaxies, including NGC~4151, NGC~3079, and CGCG~420--015~\cite{abbasi2025icecube,neronov2024neutrino}. These findings indicate that Seyfert galaxies currently represent the most prominent class of high-energy neutrino emitters.

Several physical mechanisms have been proposed to account for the acceleration of nonthermal particles in the nuclear regions of active galaxies, particularly in the context of Seyfert nuclei such as NGC~1068. These mechanisms naturally lead to a variety of theoretical scenarios aiming to explain the observed neutrino. On the smallest spatial scales, protons may be stochastically accelerated in the hot coronal plasma and subsequently interact with intense X-ray photon fields, producing neutrinos and gamma rays via hadronic processes (see e.g. ~\cite{murase2020hidden,kheirandish2021high,lemoine2025neutrinos}). Particle acceleration driven by magnetic reconnection in the corona or in turbulent plasma environments has also been extensively explored (see e.g.~\cite{mbarek2024interplay,fiorillo2024tev,fiorillo2025contribution,2025arXiv250706110S}), as well as proton acceleration at accretion shocks close to the central engine~\cite{murase2024sub}. 

On larger spatial scales, high-energy neutrino and gamma-ray production may arise from interactions between AGN-driven jets and the surrounding interstellar medium (ISM)~\cite{fang2023high}. More complex frameworks have also been proposed, such as two-zone models that combine contributions from the nuclear corona and circumnuclear starburst regions, in order to reproduce the observed gamma-ray emission~\cite{eichmann2022solving}. In addition, transient events occurring within AGN accretion disks, including supernova explosions and compact binary mergers, have been suggested as potential sources of high-energy neutrinos~\cite{zhu2021thermonuclear,ma2024high,zhou2023high,zhou2023high1}. Similar particle acceleration and neutrino production mechanisms have also been discussed in the context of jetted AGNs, most notably for the blazar TXS~0506+056~\cite{icecube2018multimessenger}.

In addition to the aforementioned models, AGNs can also drive outflows with velocities ranging from \( 300~{\rm km~s^{-1}} \) to \( 0.3c \), which are believed to be launched from the accretion disk via radiative or magnetohydrodynamic (MHD) mechanisms~\cite{laha2021ionized,harrison2018agn}. Such outflows can generate shocks in the vicinity of the AGN, where particles may be accelerated by diffusive shock acceleration (DSA)~\cite{blandford1987particle,bell2013cosmic,malkov2001nonlinear}. Several studies have invoked outflows to explain the neutrino emission observed from NGC~1068~\cite{inoue2022high,lamastra2016galactic}. In our previous study, we proposed that AGN-driven outflows may interact with dark clouds embedded in the coronal region~\cite{huang2024high}. The bow shocks generated by the outflow–cloud interactions can accelerate protons, which then undergo hadronic interactions with the ambient gas or radiation fields, producing high-energy neutrinos. This mechanism can potentially account for the neutrino emission observed from NGC~1068. In this work, we present a more detailed analysis of the outflow–cloud interaction model for Seyfert nuclei. We investigate the dominant hadronic processes and emphasize the parameter dependencies within this framework. We then extend our study to all known neutrino-associated Seyfert galaxies to reproduce their observed neutrino and gamma-ray fluxes. Finally, we explore the contributions of the entire Seyfert galaxy population to the diffuse neutrino and gamma-ray backgrounds.

The structure of this paper is as follows. In Section~\ref{sec:model}, we review the physical framework of the outflow–cloud interaction model and describe the hadronic processes involved. We apply the model to individual Seyfert galaxies, using NGC~1068 as a case study to illustrate parameter dependencies. We also evaluate the contributions to the diffuse neutrino and gamma-ray backgrounds from the overall Seyfert population. Finally, we summarize our findings in Section~\ref{sec:sum}.

\section{Model Review}\label{sec:model}
\subsection{Dynamic Process}
As discussed in previous studies of NGC~1068, the observed neutrino emission from Seyfert galaxies is likely produced in the AGN corona region~\cite{inoue2020origin,anchordoqui2021high,murase2022hidden,padovani2024high,kheirandish2021high}. 
In our scenario, we assume the presence of long-lived, clumpy gas that is distributed quasi-isotropically and uniformly around the supermassive black hole (SMBH) within the corona region. These clumps (hereafter referred to as \textit{clouds}) may originate from supernova explosions in the inner region of the star-forming disk~\cite{wang2012star}, and are subsequently driven outward along magnetic field lines by the disk radiation pressure, eventually forming a metal-rich broad-line region (BLR)~\cite{czerny2011origin,2024Univ...10...29N}. The clouds are assumed to be long-lived, as continuous mass circulation between the star-forming disk and the BLR can sustain the gas supply.Another potential formation channel is via coronal mass ejections (CMEs) from the abundant stars in the nuclear region. Although the supernovae and CMEs may occure at a relatively large distance from SMBH, a significant fraction of fragments from supernovae or CMEs can be captured by SMBH with a pericentre radius smaller than the corona radius, forming sufficient clouds in the vicinity of SMBH. A numerical calculation is implemented in Appendix~\ref{sec:cloud} to evaluate the cloud supply.

The geometric size and number density of such clouds are set to $r_{\rm c} \simeq 10^9 \  \rm {cm}$ and $n_{\rm c}\simeq5\times10^{22} \ {\rm cm^{-3}}$, based on the typical density of the outer part of red giant star~\cite{passy2012survival}. The cloud location is parameterized as \( r_0 = \mathcal{R} R_{\rm s} \), where \( R_{\rm s} = 2GM_{\bullet}/c^2 \) is the Schwarzschild radius and \( M_{\bullet} \) is the mass of SMBH. In our model,  $r_0$ is treated as a free parameter, while it must be larger than the tidal disruption radius $r_{\rm d},$ i.e., $r_0 > r_{\rm d} \simeq 13 R_{\rm s,12.5} \, M_{\bullet,7}^{-2/3} \, n_{\rm c,22.7}^{-1/3}.$ to avoid the SMBH disrupting the clouds~\cite{rossi2021process}. The conventional notation \( Q_{x} = Q / 10^x \) in cgs units is adopted hereafter.

\begin{figure}[t]
    \centering
\includegraphics[width=0.6\linewidth,height=0.5\linewidth]{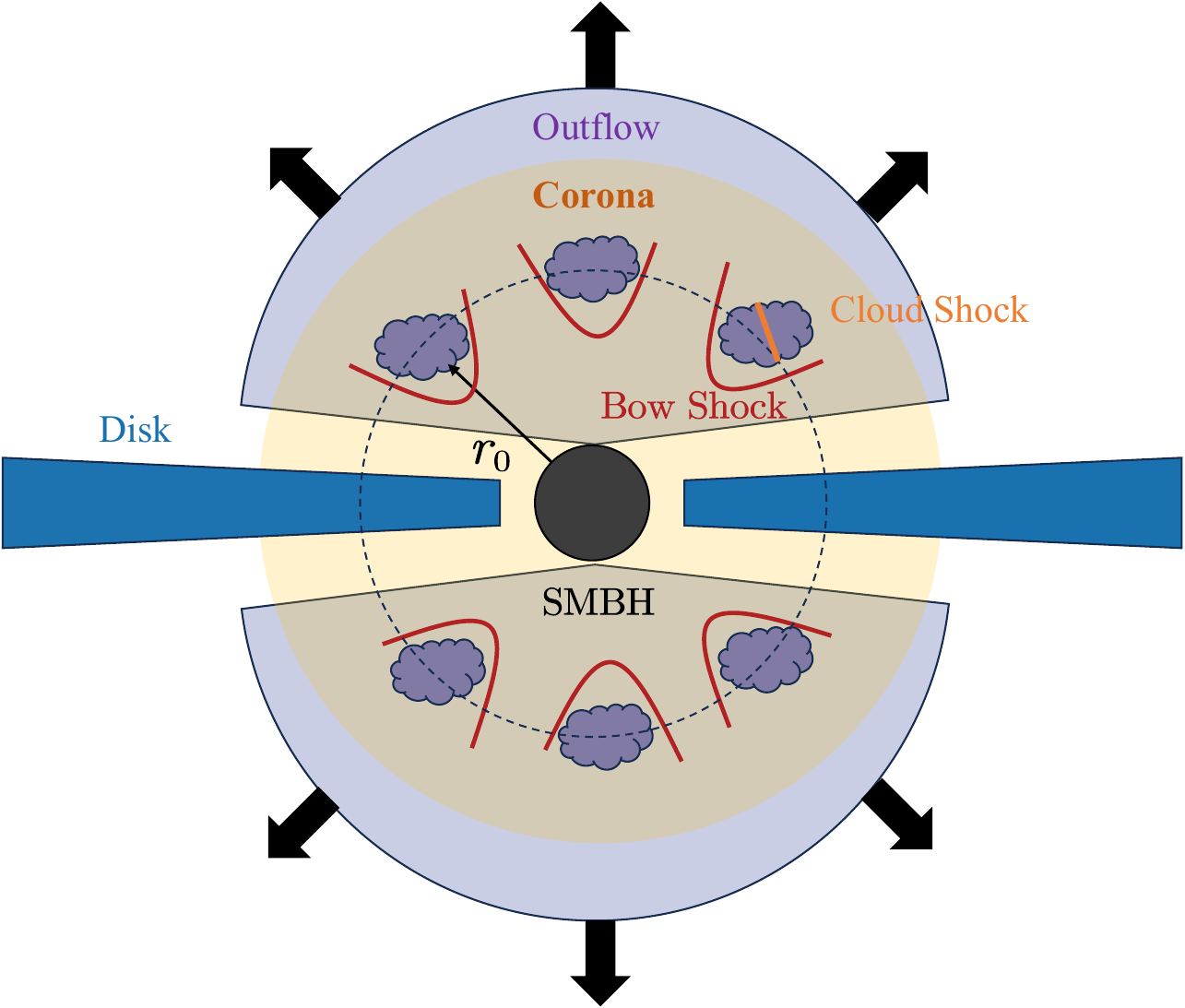}
\caption{A schematic illustration (not to scale) of the proposed outlfow-cloud intercation model.}
\label{fig:sketch}
\end{figure}

 In Seyfert galaxies, AGNs can produce quasi-isotropic outflows. Such AGN-driven winds may originate from accretion disks through multiple launching mechanisms involving thermal, radiative, and magnetic processes~\cite{crenshaw2003mass,ohsuga2014outflow}. The typical outflow velocity is in the range of \( v_0 \simeq 0.03\text{--}0.3c \)~\cite{peretti2023gamma,mizumoto2019kinetic}.  In principle, such outflows can manifest as blue-shifted atomic absorption features in the ultraviolet (UV) to X-ray bands~\cite{veilleux2005galactic,king2015powerful,laha2021ionized}. However, direct observations of outflows in Seyfert galaxies are challenging, largely due to obscuration by the thick gas and dust~\cite{garcia2016alma,gamez2022thermal,matt1997hard}. Nevertheless, there is observational evidence supporting their existence in some cases, such as the well-studied Seyfert galaxy NGC~4151~\cite{peretti2023gamma}. The associated kinetic luminosity is often parameterized as a fraction of the AGN bolometric luminosity, i.e., \( L_{\rm kin} = \eta_k L_{\rm bol} \), where $\eta_{\rm k} \leq 1$. These high-speed, quasi-isotropic outflows may collide with the clouds at a distance \( r_0 \)~\cite{wu2022could,2024PhRvD.110d3029W}, generating a bow shock outside the cloud and a cloud shock inside it, with characteristic velocities \( v_0 \) and \( v_{\rm c} \), respectively~\cite{mckee1975interaction}.
 We can connect two shock velocities through the relation $v_{\rm c}\simeq (n_0/n_{\rm c})^{1/2}v_0 $, where $n_0\simeq6.6\times10^{12} \ \eta_{\rm k,-1}L_{\rm bol,45}(\mathcal{R}/15)^{-2}R_{\rm s,12.5}^{-2}(v_0/0.03c)^{-3}\ {\rm cm}^{-3}$ is the outflow number density at $r_0$~\cite{mckee1975interaction,mou2021years}.
 First, we examine whether the shocks are radiation-dominated. For the bow shock, the optical depth for Compton scattering of electrons in the upstream is
$\tau \sim 6\times10^{-3}\, n_{0,13}\, \delta_9 \ll 1$, while for the cloud shock it is
$\tau \sim 6\times10^6\, n_{c,22}\, \delta_9 \gg 1,$
assuming the typical acceleration regions of the bow shock and cloud shock are comparable to the cloud size~\citep{barkov2010gamma,barkov2012interpretation}. This indicates that the cloud shock is radiation-dominated and therefore cannot efficiently accelerate particles, while the bow shock is not radiation-dominated. In addition, in the upstream region, the ion plasma frequency is $\omega_i \simeq 4\times10^9\, n_{0,13}\ \rm s^{-1},$ while the ion Coulomb frequency is
$\omega_{\rm Cou} \simeq 4\times10^{-4}\, n_{0,13}\, T_e^{-2} \left( v_0/0.03c\right)\rm~s^{-1},$
where we assume that the proton velocity is comparable to the shock velocity and that the proton temperature is equal to the electron temperature. Since \(\omega_{\rm Cou} \ll \omega_i\), the bow shock is dominated by plasma collective instabilities, making it plausible for the bow shock to accelerate particles via DSA processes~\cite{blandford1987particle,drury1983introduction}. Thus, we only consider the neutrino and gamma-ray production from the bow shock in this study.A sketch for our model is shown in Fig.~\ref{fig:sketch}. Particles are continuously accelerated until the cloud shock has swept through the entire cloud; the total duration of the acceleration process is thus set by~\cite{klein1994hydrodynamic}
\begin{equation}\label{eq:cloud}
    \begin{aligned}
        t_{\rm cloud}& = \frac{r_{\rm c}}{v_{\rm c}}
        \simeq 9.7 \times 10^5  \, \eta_{\rm k,-1}^{-1/2}r_{\rm c,9} 
        \left( \frac{\mathcal{R}}{15} \right) R_{\rm s,12.5}r_{\rm c,9}  
        \\
        &\quad\quad\quad\quad \quad\times 
        \left( \frac{n_{\rm c, 22.7}}{L_{\rm bol,45}} \right)^{1/2} \left( \frac{v_0}{0.03c} \right)^{1/2} \ {\rm s}.
    \end{aligned}
\end{equation}

\subsection{Hadronic Processes}
In this section, we analyze the timescales for particle acceleration and interaction to determine the dominant process responsible for neutrino production. The bow shock can efficiently accelerate protons to high energies via the DSA mechanism. The total proton luminosity is estimated as  
\( L_{\rm p} \simeq \alpha \beta L_{\rm kin} = 9\times10^{42}~\, \eta_{\rm k,-1} L_{\rm bol,45}~{\rm erg}~{\rm s}^{-1} \),  
where \( \alpha \sim 0.3 \) is the cloud covering factor and \( \beta \sim 0.3 \) is the energy fraction transferred to accelerated particles. The corresponding acceleration timescale can be estimated through~\cite{drury1983introduction}

\begin{equation}\label{eq:acc}
\begin{aligned}
    t^{-1}_{\rm acc} &= \eta_{\rm {acc}}\frac{qBv_0^2}{E_{\rm p}c} \simeq 0.06~\eta^{1/2}_{\rm k,-1} L^{1/2}_{\rm bol,45}\epsilon^{1/2}_{\rm B,-2} \left( \frac{\mathcal{R}}{15} \right)^{-1} R^{-1}_{\rm s,12.5} \\
    &\quad\quad\quad\quad\quad\quad \times \left( \frac{v_0}{0.03c} \right)^{3/2} \left( \frac{E_{\rm p}}{50~\mathrm{TeV}} \right)^{-1}~\mathrm{s}^{-1},
\end{aligned}
\end{equation}
where $\eta_{\rm acc}$ represents the efficiency of the particle acceleration process, whose value remains uncertain. In this work, we consider a broad range of $\eta_{\rm acc} = 0.01$--$1$. \( q \) is the proton charge, \( E_{\rm p} \) is the proton energy, and \( \epsilon_{\rm B} \) is the fraction of magnetic energy to the outflow kinetic energy, defined through $B^2/8\pi=\epsilon_{\rm B}\cdot L_{\rm kin}/(4\pi r^2_0v_0)$. In many studies, the magnetic field is parameterized as
$\xi_{\rm B} = U_{\rm B}/U_{\gamma}$~\cite{murase2022hidden,das2024revealing},
which differs from our definition of $\epsilon_{\rm B}$. The acceleration timescale adopted here corresponds to the case of a perpendicular magnetic field at the shock, which yields the most efficient particle acceleration and thus provides a conservative estimate of the maximum achievable particle energy. The accelerated protons will have $pp$ collisions with cold protons of the outflow, whose timescale is
\begin{equation}\label{eq:pp}
    \begin{aligned}
        t_{\rm pp}^{-1} &\approx 0.5\, \sigma_{\rm pp} n_0 c
        \simeq 3 \times 10^{-3}  \, \eta_{\rm k,-1} L_{\rm bol,45}\\ 
        &\quad\quad\quad\quad\quad\quad \times \left( \frac{\mathcal{R}}{15} \right)^{-2} R_{\rm s,12.5}^{-2} 
        \left( \frac{v_0}{0.03c} \right)^{-3}\ {\rm s}^{-1},
    \end{aligned}
\end{equation}
where a constant $pp$ cross section, $\sigma_{pp} \simeq 5\times10^{-26}\,{\rm cm}^2$, is adopted for analytical estimates~\cite{particle2004ata}, while a more accurate energy-dependent form is used in the following numerical calculations.
\begin{figure}[t]
    \centering
\includegraphics[width=0.6\linewidth,height=0.5\linewidth]{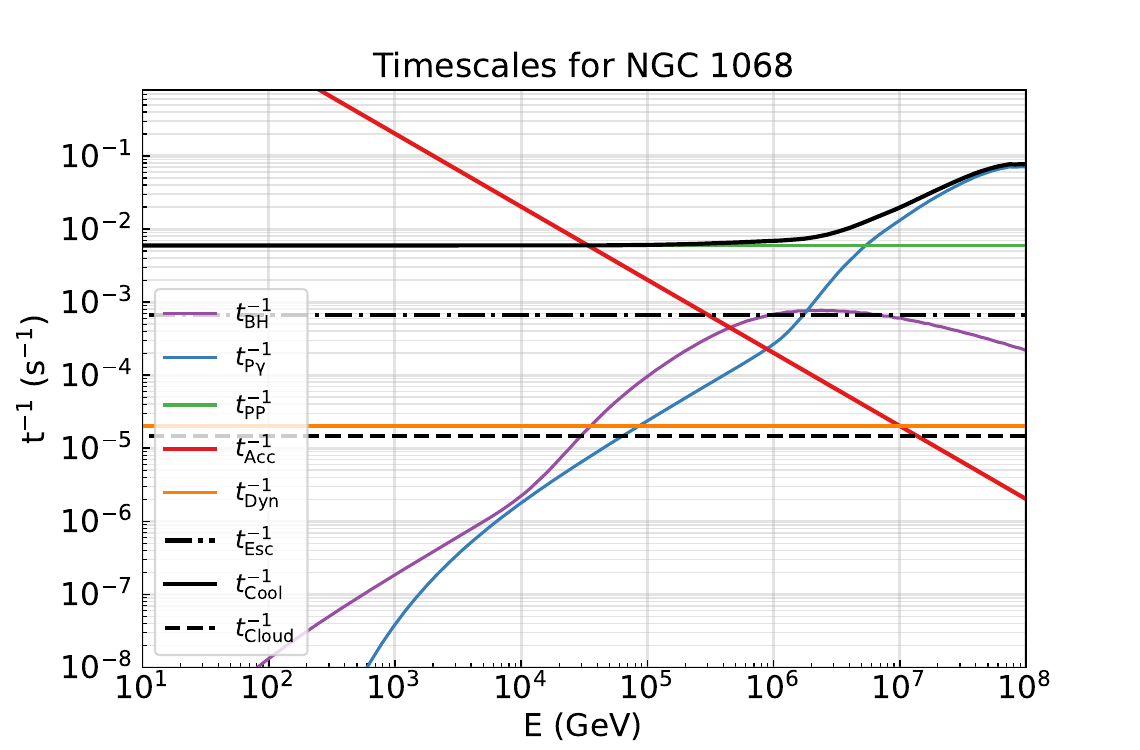}
\caption{Example proton interaction timescales for NGC 1068. Detailed expressions for the timescale calculations are provided in the main text. The adopted parameters are: $\mathcal{R} = 15$, $\epsilon_{\rm B} = 0.01$, $\eta_{\rm k} = 0.1$, and $v_0 = 0.03c$.}
\label{fig:ts}
\end{figure}

The accelerated protons can also undergo \( p\gamma \) interactions with soft photons originating from the corona and the disk. We model both photon fields following the approach proposed by Ref.~\cite{murase2020hidden}, in which the AGN bolometric luminosity \( L_{\rm bol} \) can be estimated from the intrinsic X-ray luminosity \( L_{\rm X} \) at 2-10 keV, using empirical correlations, see Eq (\ref{eq:bol}). The spectral energy distribution (SED) of AGN is parameterized as a function of the Eddington ratio~\cite{ho2008nuclear}, defined as \( \lambda_{\rm Edd} = L_{\rm bol}/L_{\rm Edd} \), where the Eddington luminosity is given by \( L_{\rm Edd} \approx 1.3 \times 10^{45}~M_{\bullet,7}~{\rm erg~s}^{-1} \). Using this method, the SED is fully determined by SMBH mass \( M_{\bullet} \) and the intrinsic X-ray luminosity \( L_{\rm X} \). See more details in Appendix~\ref{sec:sph}.

The disk spectrum has the form of multi-color blackbody~\cite{shakura1973black}, while the X-ray spectrum follows a power-law with an exponential cutoff~\cite{trakhtenbrot2017swift,ricci2018bat}, i.e. $dn_{\rm X}/d\epsilon_{\rm X} \propto n^{-\Gamma_{\rm X}}_{\rm X}\exp(-\epsilon_{\rm X}/\epsilon_{\rm X, cut})$. Based on this form of X-ray spectrum, the $p\gamma$ interaction timescale between protons and X-ray photons is~\cite{murase2008high,murase2016hidden}
\begin{equation}\label{eq:pg_x}
\begin{aligned}
    t^{-1}_{p\gamma,\rm X} 
    &\simeq \eta_{p\gamma} \, \sigma_{p\gamma} \,
    \frac{L_{\rm X}}{4\pi r_0^2 \epsilon_{\rm X}} 
    \left( \frac{E_{\rm p}}{\tilde{E}_{p\text{-}X}} \right)^{\Gamma_{\rm X} - 1} \\
    &\overset{(\Gamma_{\rm X} \approx 2)}{\simeq} 
    2.1 \times 10^{-5}~L_{{\rm X},43.7}
    \left( \frac{\mathcal{R}}{15} \right)^{-2} R_{{\rm s},12.5}^{-2} 
    \left( \frac{E_{\rm p}}{50~\mathrm{TeV}} \right)~\mathrm{s}^{-1},
\end{aligned}
\end{equation}
where $\eta_{p\gamma}=2/(1+\Gamma_{\rm X})$, $\sigma_{p\gamma}\sim0.7\times10^{-28}~{\rm cm}^{-2}$ is the cross section for photomeson interaction and $\tilde{E}_{p-X}=0.5m_{\rm p}c^2\cdot0.3~{\rm GeV}/\epsilon_{\rm X}$ is the typical proton energy interacted with photons of energy  $\epsilon_{\rm X}$~\cite{stecker1968effect}. From Eq (\ref{eq:pp}) and Eq (\ref{eq:pg_x}), we can write the timescale ratio between $pp$ collisions and $p\gamma$ interaction with corona photons as
\begin{equation}\label{eq_ratio1}
    \frac{t_{p\gamma,{\rm X}}}{t_{pp}}\simeq 72.1~\eta_{\rm k,-1}L_{\rm bol, 45}L^{-1}_{\rm X,43.7}\left( \frac{v_0}{0.03c} \right)^{-3}\left( \frac{E_{\rm p}}{50~{\rm TeV}} \right)^{-1}.
\end{equation}

The observed neutrino energies from Seyfert galaxies are in the range $E_{\nu} \sim  0.3-30 $ TeV (excluding NGC 7469, from which neutrino energies $>$ 100 TeV), implying parent proton energies of $E_{\rm p}\sim 6-600$ TeV, with a typical energy ratio of $\sim 20$ between parent protons and neutrinos produced. From Eq (\ref{eq_ratio1}), the corresponding timescale ratio in this proton energy range is $t_{p\gamma,{\rm X}}/t_{pp}\sim 6.0-600.8$, suggesting that the $p\gamma$ interaction with X-ray photons is much less efficient than the $pp$ collision in producing neutrinos. Similarly, for the $p\gamma$ process between protons and disk photons, we have
\begin{equation}\label{eq:disk}
    \begin{aligned}
        t^{-1}_{p\gamma,{\rm d}}&\simeq\frac{\sigma_{p\gamma}L_{\rm bol}}{4\pi r^2_0\epsilon_{\rm d}}\\
        &\simeq 0.06~L_{\rm bol,45}\left( \frac{\mathcal{R}}{15} \right)^{-2} R_{\rm s,12.5}^{-2}\left( \frac{\epsilon_{\rm d}}{30~{\rm eV}} \right)^{-1}~{\rm s}^{-1},
    \end{aligned}
\end{equation}
where the disk luminosity is approximated by the bolometric luminosity $L_{\rm bol}$ and $\epsilon_{\rm d}$ is the maximum photon energy of the disk~\cite{woo2002active}, which can be estimated by the effective
temperature at the innermost stable circular orbit (ISCO). We can also evaluate the timescale ratio between $p\gamma$ interaction with the disk photon and the $pp$ collisions as 
\begin{equation}
    \frac{t_{p\gamma,{\rm d}}}{t_{pp}}\simeq 0.06~\eta_{\rm k,-1}\left( \frac{v_0}{0.03c} \right)^{-3}\left( \frac{\epsilon_{\rm d}}{30~{\rm eV}} \right)^{-1},
\end{equation}
which suggests that this process could dominate over $pp$ collisions. However, the typical proton energy required for this interaction is $\tilde{E}_{p-X}=0.5m_{\rm p}c^2\cdot0.3~{\rm GeV}/\epsilon_{\rm d}\simeq 5~{\rm PeV}$. The corresponding neutrino energy is \(\sim250~{\rm TeV}\), which is much greater than most of the observed neutrino energies, implying that $p\gamma$ may contribute to the highest end of the neutrino spectrum but is unlikely to account for the entire observed energy range. Therefore, \(pp\) collisions are expected to be the dominant mechanism for neutrino production in this model.

Protons may lose energy through the Bethe–Heitler (BH) process at lower energies , which could suppress neutrino production in the low-energy band. The maximum efficiency of this process occurs during interactions with disk photons~\cite{murase2022hidden}, for which the timescale is  
\begin{equation}\label{eq:BH}
    \begin{aligned}
        t^{-1}_{{\rm BH},{\rm d}} &\simeq \frac{\sigma_{\rm BH}L_{\rm bol}}{4\pi r^2_0\epsilon_{\rm d}}\\
        &\simeq 6\times10^{-4}~L_{\rm bol,45}\left( \frac{\mathcal{R}}{15} \right)^{-2} R_{\rm s,12.5}^{-2}\left( \frac{\epsilon_{\rm d}}{30~{\rm eV}} \right)~{\rm s}^{-1},
    \end{aligned}
\end{equation}
where BH cross section is taken as \(\sigma_{\rm BH} \sim 0.8\times10^{-30}~{\rm cm^{-2}}\). The typical proton energy is  
\(\tilde{E}_{\rm BH-X} = 0.5m_{\rm p}c^2 \cdot 10~{\rm MeV}/\epsilon_{\rm d} \simeq 156.6~{\rm TeV}\)~\cite{chodorowski1992reaction,stepney1983numerical}.  
At this proton energy, the corresponding \(p\gamma\) process is dominated by interactions with X-ray photons. We find  
\(t_{\rm BH,{\rm d}}/t_{p\gamma,{\rm X}} \approx 0.11\) and \(t_{\rm BH, {\rm d}}/t_{pp} \approx 5\),  
suggesting that although the BH process dominates over the \(p\gamma\) channel in the low-energy regime, it remains less efficient than \(pp\) collisions. Therefore, the overall suppression of neutrino production due to the BH process should be limited, in contrast to the results in Ref.~\cite{inoue2020origin,murase2020hidden}, where BH-induced suppression was considered significant because only the \( p\gamma \) channel was taken into account in their scenarios.

We define the total proton cooling timescale as $t^{-1}_{\rm cool}=t^{-1}_{pp}+t^{-1}_{p\gamma}+t^{-1}_{\rm BH}$, where $t_{p\gamma}$ and $t_{\rm BH}$ include contributions from both disk and X-ray photons. The maximum proton energy, $E_{\rm p,max}$, can then be determined via $t_{\rm acc}=\min \{ t_{\rm cool},t_{\rm cloud},t_{\rm esc}\}$, where ballistic escape timescale $t_{\rm esc}=R/c$.  From Eq (\ref{eq:pp})- Eq (\ref{eq:BH}), we typically find that $t_{\rm cool}/t_{\rm cloud/esc} \ll 1$, indicating that $E_{\rm p,max}$ is generally constrained by radiative cooling rather than the cloud's dynamical lifetime. Assuming the proton radiative cooling is dominated by $pp$ collisions, we can combine Eq (\ref{eq:acc}) and (\ref{eq:pp}) to write the formula for $E_{\rm p,max}$ as
\begin{equation}\label{eq:Epmax}
    \begin{aligned}
        E_{\rm p,max}&\simeq 1~\left( \frac{\epsilon_{\rm B,-2}}{\eta_{\rm k,-1}}\right)^{1/2}L^{-1/2}_{\rm bol,45}\left(\frac{\mathcal{R}}{15}\right)R_{\rm s,12.5}\left(\frac{v_0}{0.03c}\right)^{9/2}~{\rm PeV}.
    \end{aligned}
\end{equation}The example proton interaction timescales for NGC 1068 are illustrated in Fig.\ref{fig:ts}, and show good agreement with the above analytical expectations. The detailed timescale for the $p\gamma$ is calculated via
\begin{equation}
t^{-1}_{p\gamma} = \frac{c}{2\gamma_p^2} \int_{\tilde{E}_{\rm th}}^{\infty} d\tilde{E} \, \sigma_{p\gamma}(\tilde{E}) \kappa_{p\gamma}(\tilde{E}) \tilde{E} \int_{\tilde{E}/2\gamma_p}^{\infty} d\epsilon \, \epsilon^{-2} \frac{dn_{\{\rm X,d\}}}{d\epsilon}, \,
\end{equation}
where $ \sigma_{p\gamma} $ and $ \kappa_{p\gamma} $ are the cross-section and inelasticity, respectively~\cite{stecker1968effect,patrignani2016review}, $ \tilde{E} $ is the photon energy in the proton rest frame, and $ \tilde{E}_{\rm th} \simeq 145~\mathrm{MeV} $ is the threshold energy. Here, $ \gamma_p = E_p / (m_pc^2) $ is the proton Lorentz factor, and \( dn_{\{\rm X,d\}}/d\epsilon \) is the differential number density of soft photons from corona or disk.  The BH process timescale $ t_{\rm BH} $ is calculated using the same expression, with $ \sigma_{p\gamma} $ and $ \kappa_{p\gamma} $ replaced by the corresponding quantities $ \sigma_{\rm BH} $ and $ \kappa_{\rm BH} $~\cite{chodorowski1992reaction}. 
With the timescales, we can estimate the neutrino flux for NGC 1068 as~\cite{murase2016hidden,wang2009prompt}
\begin{equation}
    \begin{aligned}
        E^2_{\nu}\frac{dN_{\nu}}{dE_{\nu}}&\simeq\frac{1}{4\pi d^2_{\rm L}}\frac{3K}{4(1+K)}f_{pp}A_{\rm n}L_{\rm p}\\
   &\overset{(\Gamma_{\rm p} \approx 2)}{\simeq} 1\times10^{-8}~\eta_{\rm k,-1}L_{\rm bol,45}~{\rm GeV}~{\rm cm^{-2}}~{\rm s}^{-1}.
    \end{aligned}
\end{equation}
which is comparable to the observed flux. Here, \( d_{\rm L} = 14.4~{\rm Mpc} \) is the luminosity distance of NGC~1068~\cite{icecube2022evidence}, and \( K = 2 \) is adopted for the $pp$-dominated case~\cite{murase2016hidden}. The parameter \( A_{\rm n} \) denotes the normalization of the proton spectrum. The factor \( f_{pp} \) represents the efficiency of $pp$ collisions, defined as \( f_{pp} = \min\{1, t_{\rm dyn}/t_{pp}\} \), where \( t_{\rm dyn} = r_0 / v_0 \) is the dynamical timescale.

In summary, within our model, the dominant hadronic channel responsible for the observed neutrinos is the $pp$ collision between accelerated protons and cold protons from the outflow. The $p\gamma$ process involving disk photons may contribute marginally, but only at the highest neutrino energies. The suppression caused by the BH process is expected to be limited.
\begin{figure*}[t]
\includegraphics[width=0.33\linewidth,height=0.3\linewidth]{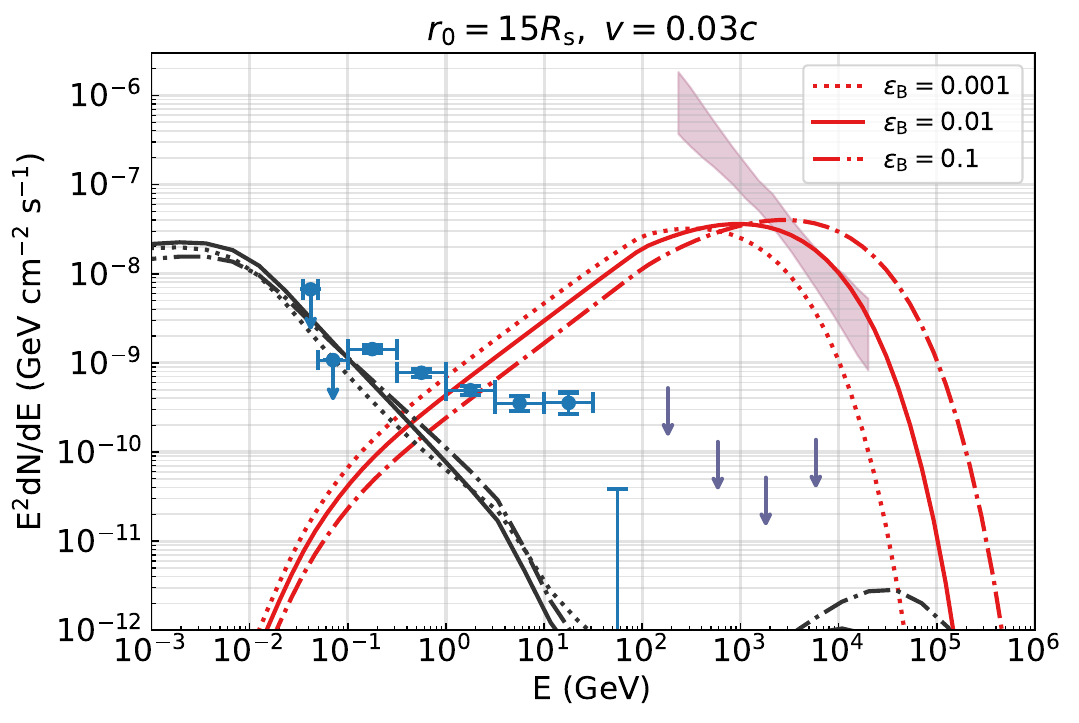}
\includegraphics[width=0.33\linewidth,height=0.3\linewidth]{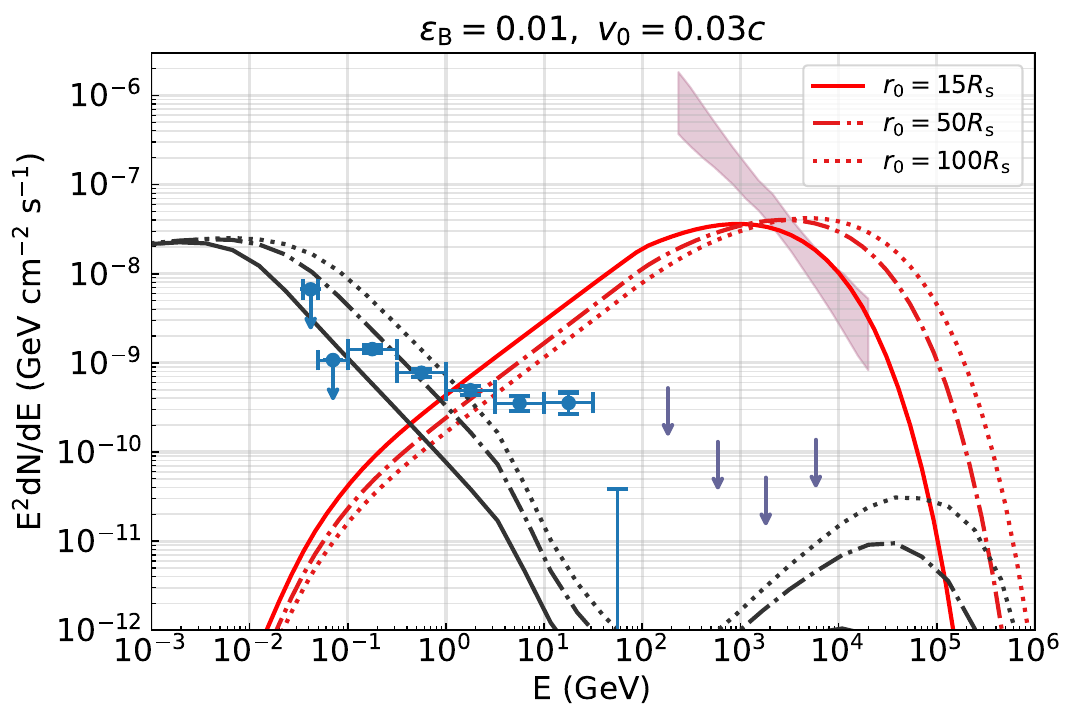}
\includegraphics[width=0.33\linewidth,height=0.3\linewidth]{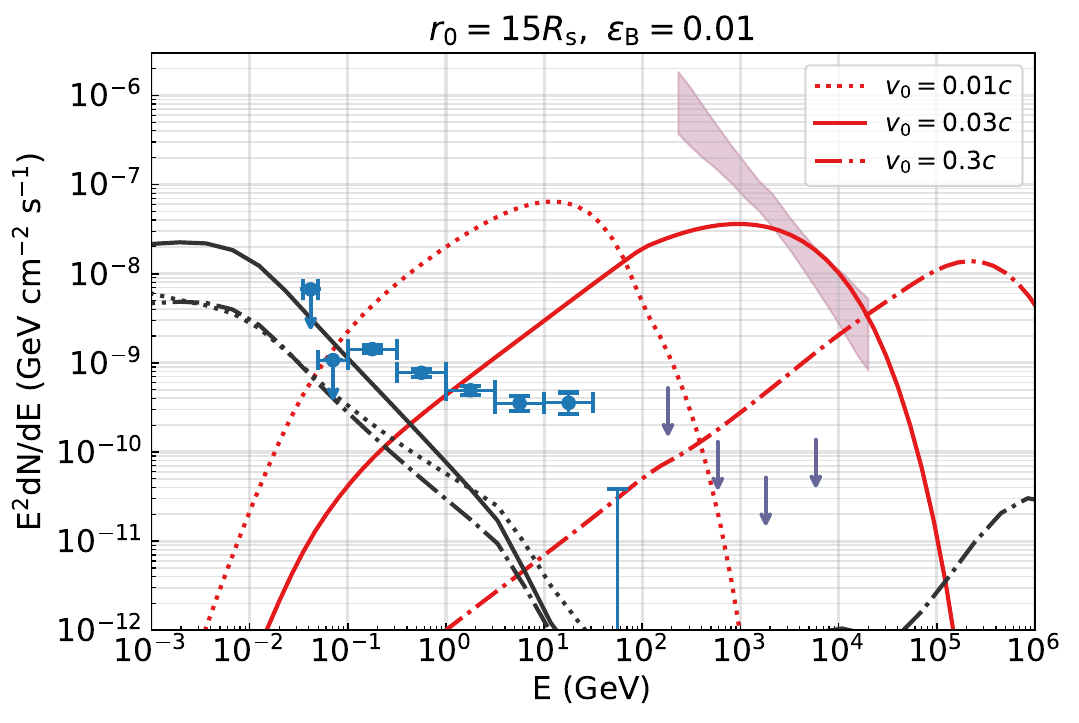}
\caption{The all-flavor neutrino and gamma-ray fluxes for NGC 1068 under various parameter combinations, with the proton spectral index fixed at $\Gamma_{\rm p} = 1.2$, energy conversion efficiency $\eta_{\rm k} = 0.2$, and acceleration efficiency $\eta_{\rm acc} = 0.02$. Red lines (with different linestyles) represent the predicted neutrino fluxes, while black lines denote the corresponding cascade photon fluxes. Blue data points show Fermi-LAT 16-year observations, and purple arrows indicate MAGIC upper limits~\cite{acciari2019constraints}. The pink shaded region indicates the neutrino flux detected by IceCube~\cite{abbasi2025icecube}, showing the 95\% confidence uncertainties and the sensitive energy range of the power-law fit for NGC~1068.}
\label{fig:compare} 
\end{figure*}

\subsection{Individual Sources}

In this section, we numerically calculate the neutrino and gamma-ray spectra of neutrino-detected Seyfert galaxies. We consider only primary protons accelerated at the bow shock and neglect primary electrons, as the focus of this work is on neutrino emission. The contribution from primary electrons is therefore assumed to be subdominant. Overall, we take the cloud parameters \( n_{\rm c} = 5\times10^{22}~{\rm cm^{-3}} \) and \( r_{\rm c} = 10^{9}~{\rm cm} \) to be constant, while treating \( \mathcal{R} \), \( \epsilon_{\rm B} \), \( \eta_{\rm k} \), and \( v_0 \) as free parameters. We assume the accelerated protons follow a single power-law distribution with an exponential cutoff~\cite{blandford1987particle}, normalized by the outflow kinetic energy density at \( r_0 \)
\begin{equation}
    \int_{1~\mathrm{GeV}}^{\infty} A_{\rm n} E_{\rm p}^{1 - \Gamma_{\rm p}} \exp\left(-\frac{E_{\rm p}}{E_{\rm p, max}}\right) dE_{\rm p} = \frac{\alpha\beta L_{\rm kin}}{2\pi r_0^2 v_0}.
\end{equation}
 We calculate neutrino, gamma-ray, and electron (positron) productions using the methods of Ref.~\cite{kelner2006energy} and Ref.~\cite{kelner2008energy} for \( pp \) collisions and \( p\gamma \) interactions, respectively. To account for proton energy losses, a suppression factor \( f_{\rm c} = \min\{1, t_{\rm cool}/t_{\rm dyn}\} \) is applied to the production spectrum to rescale the neutrino yield. The corona and disk radiation fields attenuate the gamma-rays produced alongside neutrinos via \(\gamma\gamma\) pair production. The optical depth due to the corona is estimated as \cite{murase2020hidden}
\begin{equation}\label{eq:tau}
    \begin{aligned}
        \tau_{\gamma\gamma,X} &\simeq 0.1 \sigma_{\rm T} \frac{L_{\rm X}}{4\pi r_0 \epsilon_{\rm X}} 
        \left( \frac{E_{\gamma}}{\tilde{E}_{\gamma\text{-}{\rm X}}} \right)^{\Gamma_{\rm X} - 1} \\
        &\overset{(\Gamma_{\rm X} \approx 2)}{\simeq} 
        472~L_{{\rm X},43.7}
        \left( \frac{\mathcal{R}}{15} \right)^{-1} R_{{\rm s},12.5}^{-1} 
        \left( \frac{E_{\gamma}}{1~\mathrm{GeV}} \right),
    \end{aligned}
\end{equation}
where $E_{\gamma}$ is the gamam-ray photon energy, $\sigma_{\rm T}$ is the Thomson scattering cross section, \( \tilde{E}_{\gamma\text{-}{\rm X}} \approx m_e^2 c^4 / \epsilon_{\rm X} \) is the typical gamma-ray photon energy. $\tau_{\gamma\gamma,{\rm X}}({\rm 1~GeV}) \gg 1 $ indicates that the initial gamma-rays are expected to cascade down to sub-GeV energies, as previously discussed in Ref.~\cite{murase2020hidden, murase2022hidden, das2024revealing}. We adopt the method of Ref.~\cite{bottcher2013leptonic} to compute the steady-state electron (positron) spectrum resulting from cascading. Both synchrotron emission and inverse Compton scattering with soft photon fields are considered to derive the final photon spectrum. Detailed description for this method is presented in Appendix~\ref{sec:csc}. For numerical calculations, the optical depth for gamma rays is evaluated as ~\cite{inoue2019high}
\begin{equation}
    \begin{aligned}
    \tau_{\gamma\gamma} = \int_{-1}^{1} d\mu \int_{\epsilon_{\rm th}}^{\infty} d\epsilon \, \frac{1 - \mu}{2} \, \frac{dn_{\{\rm X,d\}}}{d\epsilon} \, \sigma_{\gamma\gamma}(E_{\gamma}, \epsilon, \theta) \, r_0,
    \end{aligned}
\end{equation}
where \( \mu = \cos\theta \), \( \epsilon_{\rm th} = \frac{2 m_e^2 c^4}{E_{\gamma}(1 - \mu)} \) is the pair production threshold energy. The pair production cross section is given by~\cite{breit1934collision, heitler1954quantum}

\begin{equation}
\begin{aligned}
\sigma_{\gamma\gamma}(E_{\gamma}, \epsilon, \theta)
&= \frac{3\sigma_T}{16} (1 - \beta^2) \\
&\quad \times \left[
2\beta(\beta^2 - 2)
+ (3 - \beta^4)
\ln\!\left(\frac{1 + \beta}{1 - \beta}\right)
\right].
\end{aligned}
\end{equation}
where $\beta = \sqrt{1 - \frac{2 m_e^2 c^4}{E_{\gamma} \epsilon (1 - \mu)}}$.

\begin{table*}[htbp]
\centering
\begin{threeparttable}
\caption{Model parameters corresponding to the numerical results shown in Fig.~\ref{fig:results}. The SMBH mass ($M_{\bullet}$) and intrinsic X-ray luminosity $L_{\rm X}$ (in the 2–10~keV band) are adopted from the literature; references are listed in the notes. The last five columns correspond to the model’s free parameters.}

\label{tab:para}
\begin{tabular}{lcccccccc}
\toprule
Source Name & $M_{\bullet}~(M_{\odot})$ &$L_{\rm X}~({\rm erg~s^{-1}})$& $v_0$~(c) & $\mathcal{R}$ & $\epsilon_{\rm B}$ & $\eta_{\rm k}$& $\Gamma_{\rm p}$& $\eta_{\rm acc}$  \\
\midrule
NGC 1068\tnote{a}   & $1.0\times10^7$&$7.0\times10^{43}$ & 0.030  & 15 &1.2  & 0.01  & 0.2 & 0.02  \\
NGC 7469\tnote{e}   &$1.0\times10^7$&$2.3\times10^{43}$ & 0.300& 15  &2.0 & 0.01 & 0.4 & 0.38 \\
NGC 4151\tnote{b}  & $1.0\times10^7$&$2.6\times10^{42}$ & 0.030 & 14  &1.2 & 0.01 & 0.4 &0.38 \\
NGC 3079\tnote{c}  & $2.0\times10^6$&$1.0\times10^{42}$ & 0.015 & 40  &2.0 & 0.01 & 0.1 &0.38 \\
CGCG 420-015\tnote{d}   &$2.0\times10^8$&$7.0\times10^{43}$ & 0.020 & 10 &1.2 & 0.01 & 0.1 &0.38 \\

\bottomrule
\end{tabular}
\begin{tablenotes}
\footnotesize
\item[a] \cite{woo2002active,marinucci2015nustar}
\item[b] \cite{bentz2022broad,koss2022bass}
\item[d] \cite{kondratko2005evidence,iyomoto2001bepposax}
\item[d] \cite{koss2017bat,tanimoto2018suzaku}
\item[e] \cite{peterson2014reverberation,ricci2021hard}
\end{tablenotes}
\end{threeparttable}
\end{table*}

\begin{figure*}[htb]
\centering
\includegraphics[width=0.32\linewidth,height=0.28\linewidth]{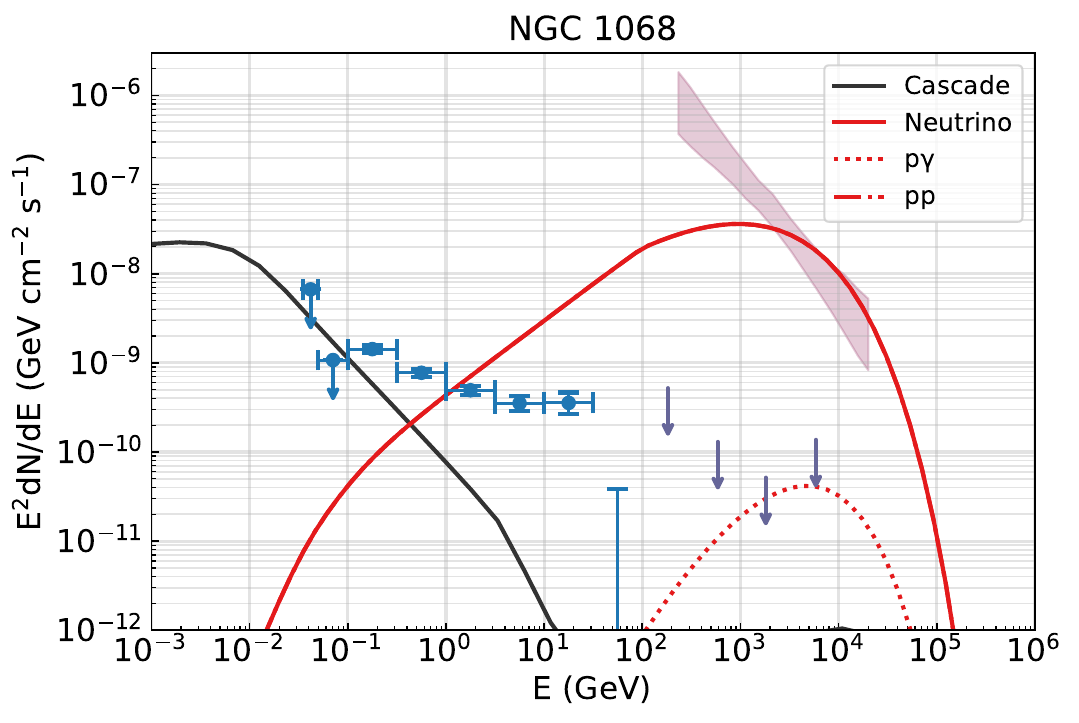}
\includegraphics[width=0.32\linewidth,height=0.28\linewidth]{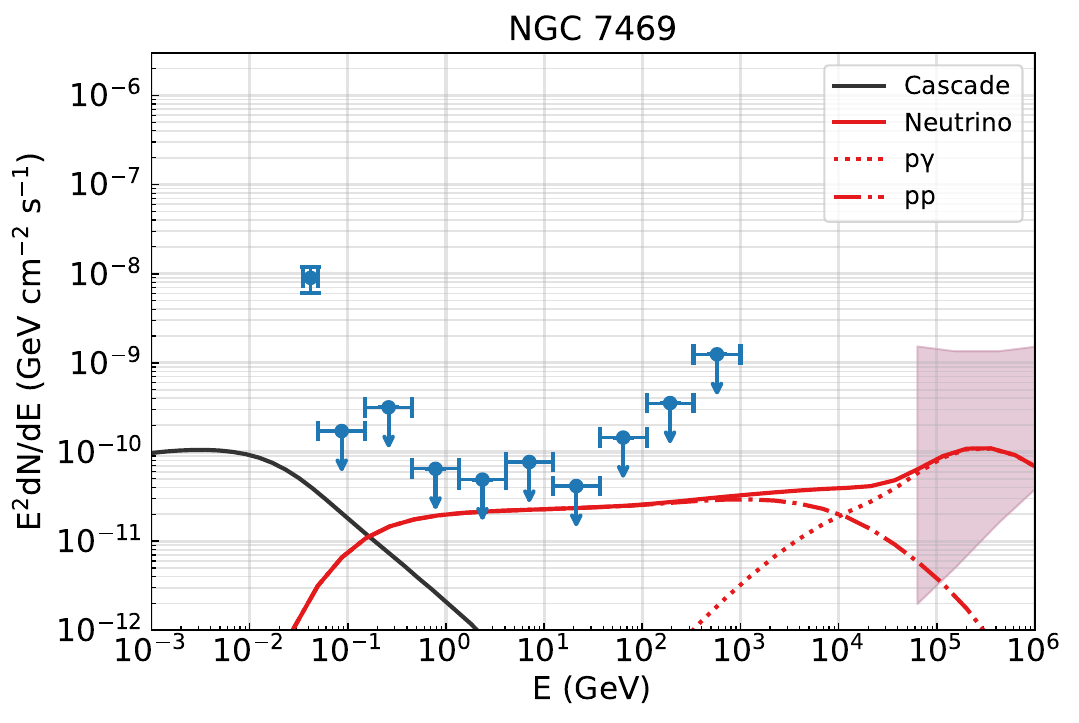}
\includegraphics[width=0.32\linewidth,height=0.28\linewidth]{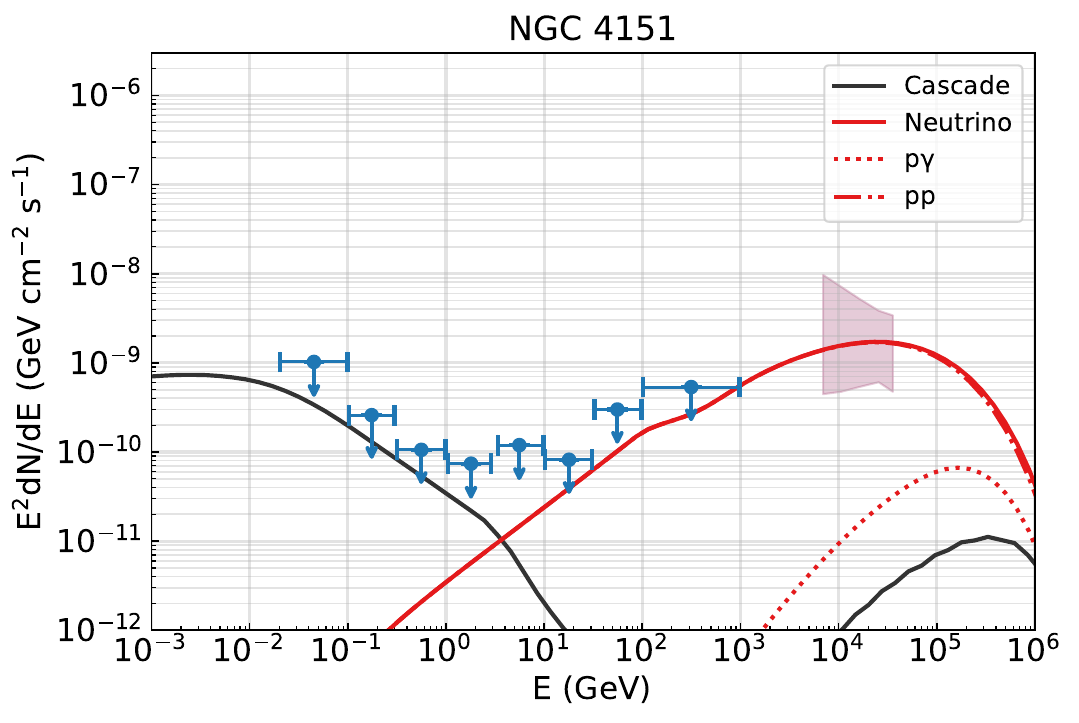}

\vspace{0.3cm}
\includegraphics[width=0.32\linewidth,height=0.28\linewidth]{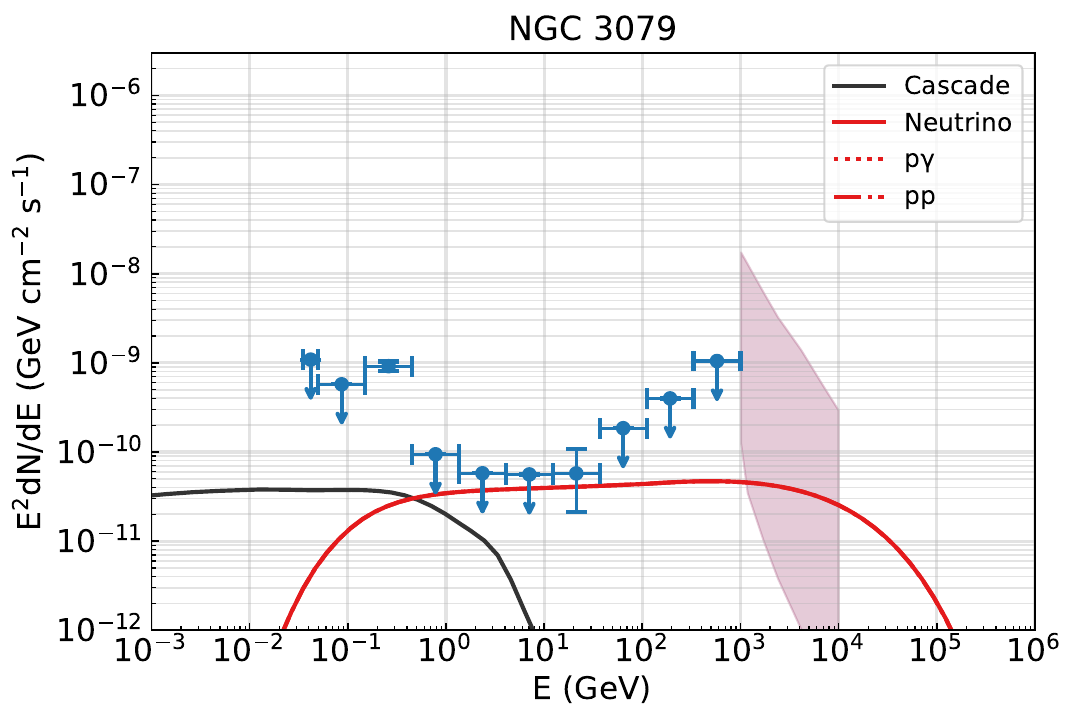}
\hspace*{0.01\linewidth}
\includegraphics[width=0.32\linewidth,height=0.28\linewidth]{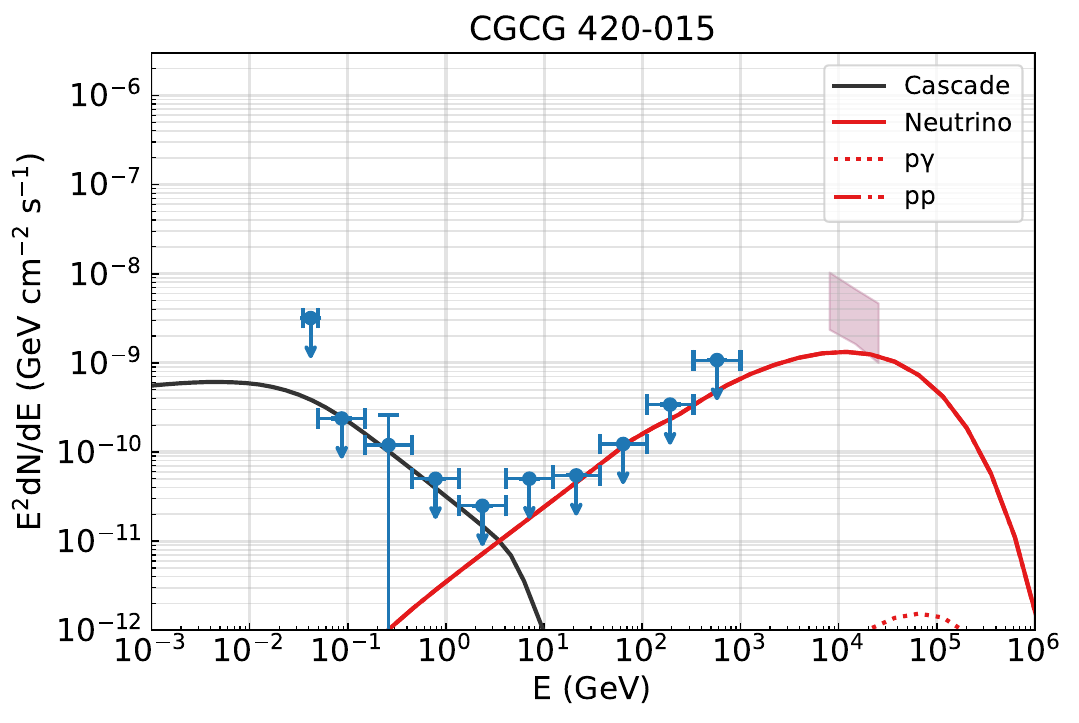}

\caption{The all-flavor neutrino and gamma-ray SEDs for five neutrino-associated sources: NGC~1068, NGC~4151, NGC~3079, CGCG~420-015, and NGC~7469. Blue points represent the 16-year Fermi-LAT gamma-ray data. The shaded regions indicate the neutrino fluxes detected by IceCube, representing the sensitive energy range of the power-law fit, with data sourced from Ref.~\cite{abbasi2025icecube} for NGC~7469, NGC~4151, and CGCG~420-015, and from Ref.~\cite{neronov2024neutrino} for NGC~3079.}
\label{fig:results}
\end{figure*}
IceCube has identified several Seyfert galaxies as potential neutrino sources, including NGC~1068, NGC~7469, NGC~4151, NGC~3079, and CGCG~420-015. The SED of soft radiation fields for these five nuclei is shown in Appendix~\ref{sec:sph}. As noted in previous studies, gamma-ray emissions in the Fermi-LAT energy band provide strong constraints on electromagnetic cascades, thereby limiting the range of viable model parameters~\cite{das2024revealing,murase2024sub}. To investigate these constraints, we analyze 16 years of Fermi-LAT data to derive the gamma-ray fluxes and 95\% C.L. upper limits for these five sources. The detailed processing is in Appendix~\ref {sec:data}. We then apply our theoretical model and compare the numerical predictions with the observed fluxes. 

Fig.~\ref{fig:compare} presents the SEDs from our model for NGC 1068 under various parameter configurations. The fiducial model adopts $\mathcal{R} = 15$, $\epsilon_{\rm B} = 0.01$, $\eta_{\rm k} = 0.2$, $\eta_{\rm acc} = 0.02$, $v_0 = 0.03c$, and a proton spectral index of $\Gamma_{\rm p} = 1.2$. To investigate the impact of physical parameters, we vary $\mathcal{R}$, $\epsilon_{\rm B}$ and $v_0$ individually while keeping the other parameters fixed. From Fig.~\ref{fig:compare}, we observe that the magnetic parameter $\epsilon_{\rm B}$ affects the maximum energy of neutrinos, but has a negligible impact on the overall flux amplitude. This is because $\epsilon_{\rm B}$ only weakly influences the maximum proton energy, following the relation $E_{\rm p,max} \propto \epsilon^{1/2}_{\rm B}$,  from Eq~(\ref{eq:Epmax}). The corresponding neutrino flux amplitude is thus affected by the normalization factor, which decreases minimally. Similar rules can be found in the case of parameter $\mathcal{R}$, the maximum neutrino energy scales with $r_0$, with relation $E_{\rm p,max}\propto r_0$ from Eq~(\ref{eq:Epmax}), while the overall flux amplitude remains nearly unchanged. The independence of the neutrino flux from \( r_0 \) can also be understood as follows: the final neutrino spectrum is normalized by the kinetic luminosity \( L_{\rm kin} \), which is independent of \( r_0 \). The parameter $v_0$ has the most pronounced impact on the resulting neutrino flux, primarily because it strongly affects the $pp$ interaction efficiency, as indicated by Eq~(\ref{eq:pp}), where $t_{\rm pp} \propto v_0^3$. At high velocities (e.g., $v_0 = 0.3c$), the $p\gamma$ interactions with disk photons begin to dominate neutrino production at the highest energies, resulting in a different shape in the spectrum, as shown in the $v_0 = 0.3c$ case in Fig.~\ref{fig:compare}.

We then extend our analysis to additional neutrino-associated Seyfert galaxies.
The numerical results, together with the observed fluxes, are shown in Fig.~\ref{fig:results}, and the corresponding model parameters for each source are summarized in Table~\ref{tab:para}. We find that NGC~3079 can be well explained by the model with a proton spectral index of $\Gamma_{\rm p} = 2$. In contrast, for NGC~1068, NGC~4151 and CGCG~420-015, a harder proton index is required to avoid cascade emission exceeding the $95\%$ C.L. upper limits. NGC~7469 is the most exceptional case, as two neutrinos with energy $> 100$~TeV  have been detected from this source, significantly more energetic than neutrinos from the others. However, our model can naturally account for the neutrino flux in this high-energy band, as \(p\gamma\) interactions with disk photons can contribute at these energies. This requires a fast outflow velocity in the source, specifically \(v_0 = 0.3c\). In most cases, the required outflow velocities range from $0.01c$ to $0.03c$, and the kinetic-to-bolometric luminosity ratio $\eta_{\rm k}$ lies between 0.1 and 0.4.

\subsection{Diffuse Neutrinos and Gamma Rays}

\begin{figure}[h]
    \centering
\includegraphics[width=0.6\linewidth,height=0.5\linewidth]{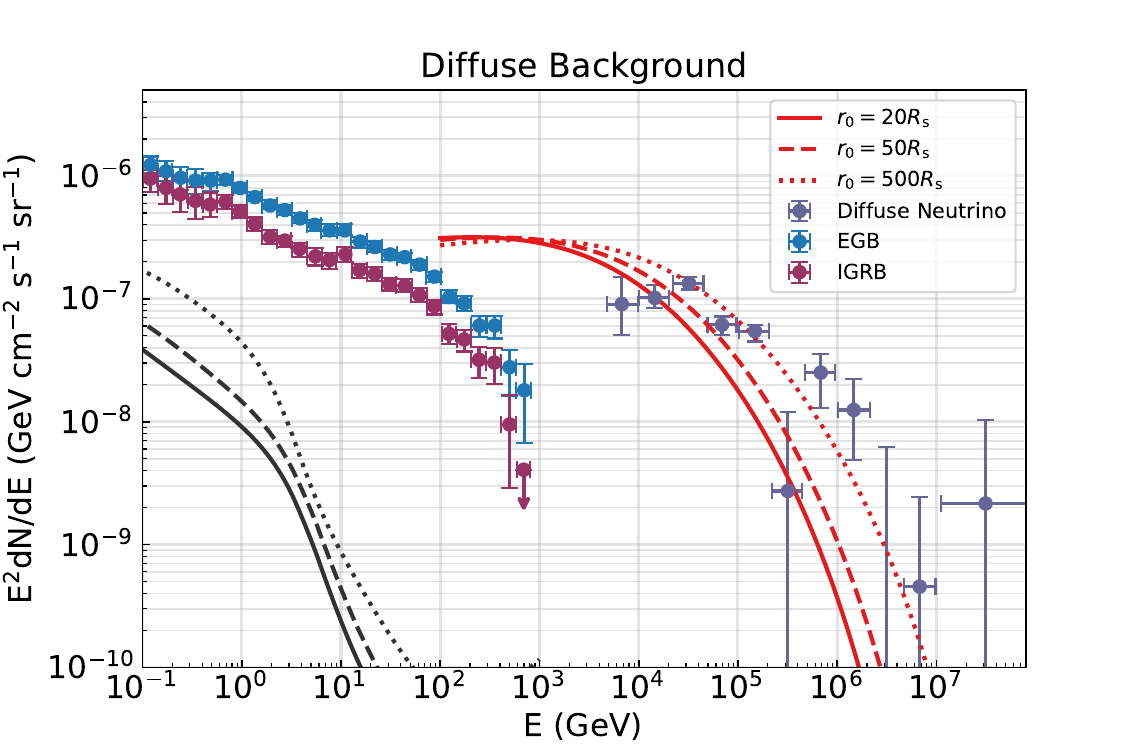}
\caption{
Diffuse neutrino and gamma-ray contributions from a population of Seyfert galaxies, assuming the Schwarzschild radius of $R_{\rm s} = 10^{12.5}~\mathrm{cm}$, which corresponds to an SMBH with mass $M_{\rm BH} = 10^7~M_{\odot}$. The extragalactic gamma-ray background (EGB) and isotropic gamma-ray background (IGRB) data are taken from Ref.~\cite{ackermann2015spectrum}, while the diffuse neutrino background data are adopted from Ref.~\cite{naab2023measurement}.
}
\label{fig:diff}
\end{figure}

The cumulative contribution of Seyfert galaxies to the diffuse neutrino background has also been explored in several works~\cite{padovani2024neutrino,murase2020hidden,fiorillo2025contribution}. While such AGNs are promising candidates, the overall contribution remains uncertain due to the diversity of source environments and model parameters.
Here, we also extend our study to the diffuse neutrino and gamma-ray fluxes originating from a population of Seyfert galaxies. The diffuse neutrino flux can be calculated as~\cite{murase2014diffuse,liu2018can,fiorillo2025contribution}
\begin{equation}
    \begin{aligned}
        E^2_{\nu} \Phi_{\nu}(E_{\nu}) &= \frac{c}{4\pi H_0} \int_0^{z_{\rm max}} \frac{dz}{\sqrt{(1+z)^3 \Omega_{\rm M} + \Omega_{\Lambda}}} \\
        &\quad \times \int d\log L_{\rm X} \, \frac{d\Psi}{d\log L_{\rm X}} \, \frac{L_{\nu}[(1+z)E_{\nu}]}{(1+z)^2},
    \end{aligned}
\end{equation}
where \( H_0 = 67.8~\mathrm{km\,s^{-1}\,Mpc^{-1}} \) is the Hubble constant, and the cosmological parameters are \( \Omega_{\rm M} = 0.308 \) and \( \Omega_{\Lambda} = 0.692 \)~\cite{ade2016planck}. Here, \( L_{\nu} \) denotes the neutrino luminosity in the source frame, and \( d\Psi/d\log L_{\rm X} \) represents the comoving number density of AGNs per logarithmic X-ray luminosity interval, as defined in Ref.~\cite{ueda2014toward}. The diffuse gamma-ray flux can be computed using the same formalism, by replacing \( L_{\nu} \) with the gamma-ray luminosity \( L_{\gamma} \). For the model parameters, we adopt $\Gamma_{\rm p} = 2$, $\epsilon_{\rm B} = 0.01$, $\eta_{\rm k} = 0.1$, and $v_0 = 0.03c$. The resulting diffuse neutrino fluxes for different values of $r_0$ are shown in Fig.~\ref{fig:diff}. We find that the diffuse neutrino emission from Seyfert galaxies can account for the observed neutrino background in the energy range of $10^4$--$10^5$~GeV.
In contrast, the associated gamma-ray contribution from cascade emissions is relatively minor, contributing only modestly in the $0.1$--$10$~GeV band.

As the cloud location parameter $r_0$ increases, the diffuse neutrino flux extends to higher energies. However, when $r_0 > 20R_{\rm s,12.5}$, the predicted flux may exceed the observed diffuse neutrino flux around $10^4$~GeV. Therefore, we constrain the typical cloud location to $r_0 < 20R_{\rm s,12.5}$ for the Seyfert galaxy population. Under this constraint, the corresponding diffuse gamma-ray flux contribution remains below 4\%.

\section{Summary}\label{sec:sum}

Seyfert galaxies are among the most promising candidates for high-energy neutrino sources. AGNs in these galaxies can drive outflows with mildly relativistic velocities that collide with dense clouds surrounding the SMBH. These outflow--cloud interactions provide natural sites for proton acceleration, leading to the production of high-energy neutrinos.

In this work, we extend the outflow-cloud interaction model to a broader population of Seyfert galaxies. We perform a detailed analysis of the hadronic processes involved and identify the dominant neutrino production channel as $pp$ collisions between accelerated protons and cold protons in the outflow. The $p\gamma$ process, involving disk photons, contributes only marginally, and only at the highest neutrino energies. Suppression due to the BH process is found to be limited. We also investigate the dependence of the model on key parameters, using NGC 1068 as a representative case.

We then apply the model to five individual Seyfert galaxies and find that their neutrino and gamma-ray emissions can be reasonably explained. However, a harder proton spectral index is required to match the observations of NGC 1068, NGC 4151, and CGCG 420--015. Finally, we estimate the diffuse neutrino and gamma-ray fluxes from the entire Seyfert galaxy population. Our results show that Seyfert galaxies can account for the observed neutrino background in the energy range of $10^4$--$10^5$~GeV. To avoid exceeding the observed diffuse flux, we constrain the typical cloud location to $r_0 < 20R_{\rm s,12.5}$. Under this condition, the corresponding diffuse gamma-ray flux contribution remains below 4\%.

Future observations by next-generation neutrino detectors such as IceCube-Gen2~\cite{aartsen2021icecube} and KM3NeT~\cite{margiotta2014km3net}, as well as upcoming MeV to sub-GeV gamma-ray missions like e-ASTROGAM~\cite{de2018science} and AMEGO~\cite{fleischhack2021amego}, will offer valuable opportunities to further test and constrain the outflow--cloud interaction scenario in Seyfert galaxies.

\appendix
\section{\label{sec:cloud} Numerical Analysis of Cloud Orbits near SMBHs}

Long-lived clouds near the SMBHs may originate from several channels. Galactic nuclear regions host abundant stars, either as part of nuclear star clusters or embedded within star-forming disks~\citep{cantiello2021stellar}. During stellar flares, these stars can generate coronal mass ejections (CMEs), which have been observed in a variety of stellar~\cite{li2026sudden,leitzinger2022stellar,ioka2020binary,lacy1976uv}. Typical CME velocities range from $10^7$ to $10^8~\rm cm~s^{-1}$, with occurrence rates of $0.01$--$10~\rm yr^{-1}$~\citep{li2026sudden}. For giant stars, individual CMEs can reach masses of $\gtrsim 10^{21}~\rm g$, occurring at rates of $\dot{R}_{\rm CME}\sim 300~\rm yr^{-1}$~\citep{argiroffi2019stellar}. These continuously injected high-velocity CMEs can be captured into stable orbits near the SMBH, thereby contributing to cloud formation.

The second similar channel is supernovae (SNe) occurring within the star-forming disk. In such events, massive stellar material can be ejected at velocities of $10^8$--$10^9~\rm cm~s^{-1}$~\citep{chevalier2005young}, and a fraction of this ejecta may settle into stable orbits near the SMBH, ultimately forming clouds in the nuclear region.

\begin{figure}[t]
    \centering
\includegraphics[width=0.6\linewidth,height=0.35\linewidth]{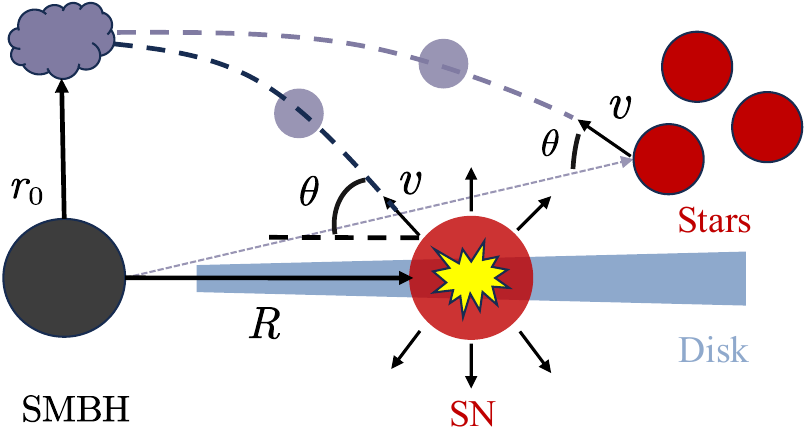}
\caption{Schematic illustration of two possible channels for dense-cloud formation near the SMBH: (i) supernova explosions within the accretion disk, and (ii) coronal mass ejections from stars.}
\label{fig:sketch_2}
\end{figure}

We argue that for both channels, even if the ejecta are launched at large radii from the SMBH ($R \gtrsim 10^{3} R_{\rm s}$), a sufficient fraction of the material can still be captured into bound orbits at small radii ($r \sim 10$--$50\,R_{\rm s}$), thereby forming the cloud region. As an illustrative example, consider an SN or a star producing CMEs at radius $R$. In the SN scenario, the explosion ejecta are assumed to be isotropic and fragment into many clumps. Each fragment is launched at an angle $\theta$ with velocity $v \sim 10^{8}\,{\rm cm\,s^{-1}}$, density $n_{\rm c} \sim 5 \times 10^{22}\,{\rm cm^{-3}}$, and size $r_{\rm c} \sim 10^{9}\,{\rm cm}$. In the CME scenario, a star ejects a single dense clump at a random angle $\theta$, with similar characteristic velocity, density, and size. The geometry of both channels is illustrated in Fig.~\ref{fig:sketch_2}.

For a single fragment of SN ejecta (or a dense clump produced by CMEs; hereafter referred to as a fragment), the specific energy and specific angular momentum are
\begin{equation}
\label{eq:orb}
    \epsilon = \frac{1}{2} v^2 - \frac{G M_{\rm eff}}{R}, \quad
    l = R v_\perp = R v \sin\theta,
\end{equation}
where $v_\perp$ is the velocity component perpendicular to the radial direction, and $M_{\rm eff}$ is the effective mass accounting for radiation pressure, defined as
\begin{equation}
    M_{\rm eff} = M_\bullet \left( 1 - \frac{\kappa L_{\rm bol}}{4 \pi G M_\bullet c} \right),
\end{equation}
with $\kappa$ being the ratio of the fragment's surface area to its mass. The orbital eccentricity and pericenter radius of the fragment are then given by
\begin{equation}
    e = \sqrt{1 + \frac{2 \epsilon l^2}{G^2 M_{\rm eff}^2}}, \quad
    r_0(R,\theta) = \frac{l^2}{G M_{\rm eff} (1 + e)}.
\end{equation}
Hence, the pericenter radius $r_0$ depends on the launch radius $R$ and angle $\theta$, which are treated as random variables. For a supernova or a star randomly distributed near the SMBH, the probability density function (PDF) of the launch radius is $g(R)$, while the launch angle $\theta$ is assumed to be uniformly distributed in $[0, \pi]$. The PDF of the pericenter radius for a single fragment is
\begin{equation}
    P(r_0) = \int_{3R_s}^{\infty} \int_0^\pi g(R) \frac{\sin\theta}{2} \, \delta\big(r_0 - r_0(R,\theta)\big) \, dR \, d\theta,
\end{equation}
where $\sin\theta/2$ arises from the uniform distribution in polar angle.

To form a stable bound orbit around the central black hole, two conditions must be satisfied: 
(1) the specific orbital energy $\epsilon$ must be negative to prevent direct escape, and 
(2) the pericenter radius $r_0$ must exceed both the Schwarzschild radius $R_s$ and the tidal radius 
\begin{equation}
r_d \simeq 13 \, R_{\rm s,12.5} \, M_{\bullet,7}^{-2/3} \, n_{\rm c,22.7}^{-1/3}
\end{equation} 
to avoid being swallowed or tidally disrupted. 

The probability for a fragment to settle into a stable orbit with pericenter $r_0$ is then
\begin{equation}
\label{eq:pdf_stable}
P(r_0 \cap \mathrm{stable}) = P(r_0 \mid \mathrm{stable}) \, P(\mathrm{stable}),
\end{equation}
where the total probability of forming a stable orbit is given by the joint condition
\begin{equation}
P(\mathrm{stable}) = P\big(\epsilon < 0,\, r_0 > \max(R_s, r_d)\big).
\end{equation}

Within these stable orbits, the probability for matter to reside in the radial interval $10$--$50\,R_{\rm s}$ can be estimated as
\begin{equation}
P_{10-50} = \int_{10R_{\rm s}}^{50R_{\rm s}} P(r_0 \cap \mathrm{stable}) \, f_{\rm orb}(r_0) \, dr_0,
\end{equation}
where $f_{\rm orb}(r_0)$ is the fractional residence time in this radial range:
\begin{equation}
f_{\rm orb}(r_0) = \frac{\int_{10R_{\rm s}}^{50R_{\rm s}} dr / v(r)}{\int_{r_0}^{r_{\rm a}} dr / v(r)},
\end{equation}
with $r_{\rm a}$ being the apocenter radius of the orbit and $v(r)$ the radial velocity, 
which can be derived from Eq.~(\ref{eq:orb}).

As an illustrative example, we adopt parameters relevant to NGC~1068: a black hole mass of $M_{\bullet}=10^{7}M_{\odot}$ and an AGN bolometric luminosity of $L_{\rm bol}=10^{45}\,{\rm erg\,s^{-1}}$. For the probability distribution of the launch radius $R$, we consider two cases. The first is a uniform distribution, in which SNe or CMEs are equally likely to occur between $3$ and $10^{6}R_{\rm s}$. The second is a Gaussian distribution in logarithmic space, with mean $\mu=3$ and standard deviation $\sigma=1$, such that most SN events are concentrated around $10^{3}R_{\rm s}$. These distributions allow us to explore how the initial location affects the fraction of ejecta captured into stable orbits.
\begin{table*}[htbp]
\begin{threeparttable}
\caption{Probabilities for a CME or a fragment of SN ejecta under different assumed probability distributions of the launch radius $R$. We adopt a black hole mass of $M_{\bullet}=10^{7} M_{\odot},L_{\rm bol}=10^{45}  {\rm erg~s^{-1}}$. Columns list the probabilities that a fragment (or CME) (i) escapes, (ii) is swallowed by the SMBH, (iii) is tidally disrupted, (iv) forms a stable bound orbit, and (v) the material actually resides within the cloud region adopted in the main text ($10$--$50\,R_{\rm s}$).
}
\label{tab:p}
\begin{tabular}{lccccc}
\toprule
$g(R)$ distribution & Escape & Swallowed & Tidal disruption  & Stable & Residence ($10$--$50~R_{\rm s}$) \\
\midrule
Uniform & $20.95\%$ & $32.20\%$ & $12.11\%$  &$34.74\%$ & $0.02\%$  \\
Gaussian & $2.46\%$ & $33.49\%$ & $23.92\%$   &$40.12\%$ & $0.03\%$  \\
\bottomrule
\end{tabular}
\end{threeparttable}
\end{table*}
The results are summarized in Table~\ref{tab:p}. We find that the probability for an individual fragment to end up on a stable orbit within the cloud region ($10$--$50\,R_{\rm s}$) is $P_{\rm 10-50}\simeq 0.02$--$0.03\%$. For SNe produced by massive stars, we assume isotropic ejecta with a total mass $M_{\rm SN}\sim10\,M_{\odot}$ and an SN rate of $\dot{R}_{\rm SN}\lesssim10^{-3}\,{\rm yr^{-1}}$ per AGN disk~\citep{grishin2021supernova}. The corresponding mass injection rate into the cloud region is therefore
\begin{equation}
\begin{aligned}
\dot{M}_{\rm SN,inj}(10\!-\!50R_{\rm s}) 
&\simeq P_{\rm 10-50} \, M_{\rm SN} \, \dot{R}_{\rm SN} \\
&\simeq (4\!-\!6)\times10^{-6}\,M_{\odot}\,{\rm yr^{-1}} .
\end{aligned}
\end{equation}

For CMEs from stars, we adopt a CME rate of $\dot{R}_{\rm CME}\sim300\,{\rm yr^{-1}}$ per star and a stellar mass density of $\rho_{\star}\sim3\times10^{7}\,M_{\odot}\,{\rm pc^{-3}}$ within $R<0.01\,{\rm pc}$, consistent with observations of $\rm Sgr~A^{*}$~\citep{schodel2018distribution}. Assuming a typical stellar mass of $M_{\star}\sim1\,M_{\odot}$, the resulting mass injection rate into the cloud region is
\begin{equation}
\begin{aligned}
\dot{M}_{\rm CME,inj}(10\!-\!50R_{\rm s})
&\simeq P_{\rm 10-50} \, M_{\rm c} \, \dot{R}_{\rm CME} \, \frac{\pi R^{3} \rho_{\star}}{M_{\star}} \\
&\simeq (1.5\!-\!2.2)\times10^{-12}\,M_{\odot}\,{\rm yr^{-1}},
\end{aligned}
\end{equation}
which is clearly negligible. In contrast, mass injection from supernovae can remain non-negligible and likely dominates the external supply.

For reference, the total cloud mass at $r\sim15\,R_{\rm s}$ is $\sim100\,M_{\odot}$ in our scenario, implying a characteristic build-up timescale of $\sim10^{7}$--$10^{8}\,{\rm yr}$. At face value, this timescale is comparable to the typical active lifetime of Seyfert galaxies, which may appear to challenge the feasibility of forming the cloud reservoir in time. However, this estimate should not be interpreted as the time required to assemble the cloud from an initially empty environment. In realistic galactic nuclei, gas is continuously supplied and recycled through stellar mass loss, supernovae, and inflows from larger scales. As a result, the cloud region is expected to be progressively built up over multiple activity cycles rather than within a single episode.

\section{\label{sec:csc}Cascade Process}
High-energy gamma rays produced through hadronic processes are attenuated by soft photons from the accretion disk or X-ray corona, resulting in the production of electron–positron pairs. These high-energy pairs subsequently emit gamma rays via synchrotron radiation and inverse Compton scattering, initiating an electromagnetic cascade that continues until a steady state is reached.
We follow the method proposed by Ref.~\cite{bottcher2013leptonic} to calculate the emission from stable electrons (positrons)  produced in the electromagnetic cascade. The steady state electron (positron) distribution, $N_{\rm e}(\gamma)$, satisfies the isotropic Fokker–Planck equation:
\begin{equation}\label{eq:FP}
    \frac{\partial}{\partial\gamma}\left(\dot{\gamma}N_{\rm e}[\gamma]\right) = Q_{\rm e}(\gamma) + \dot{N}_{\rm e}^{\gamma\gamma}(\gamma) + \dot{N}_{\rm e}^{\rm esc},
\end{equation}
where $\gamma$ is the electron (positron) Lorentz factor. The term $Q_{\rm e}(\gamma)$ represents the injection rate of electrons (positrons) from both pp ($p\gamma$) and Bethe–Heitler (BH) processes. The escape term is energy-independent and given by $\dot{N}_{\rm e}^{\rm esc} = -N_{\rm e}(\gamma)/t_{\rm esc}$, where the escape timescale is $t_{\rm esc} = r_0/c$. The total energy loss rate, $\dot{\gamma} = \dot{\gamma}_{\rm syn} + \dot{\gamma}_{\rm Com}$, includes contributions from both synchrotron radiation and inverse Compton scattering with soft photon fields (from the disk and corona). The term $\dot{N}_{\rm e}^{\gamma\gamma}(\gamma)$ denotes the injection rate of electrons (positrons) due to $\gamma\gamma$ absorption, which is given by
\begin{equation}
\begin{aligned}
     \dot{N}_{\rm e}^{\gamma\gamma}(\gamma) &= f_{\rm abs}(\epsilon_1) \left( \dot{N}^0_{\epsilon_1} + \dot{N}^{\rm syn}_{\epsilon_1} + \dot{N}^{\rm Com}_{\epsilon_1} \right)\\ 
     &+ f_{\rm abs}(\epsilon_2) \left( \dot{N}^0_{\epsilon_2} + \dot{N}^{\rm syn}_{\epsilon_2} + \dot{N}^{\rm Com}_{\epsilon_2} \right),   
\end{aligned}
\end{equation}
where the energies of the absorbed high-energy photons are $\epsilon_1 = \gamma / f_{\gamma}$ and $\epsilon_2 = \gamma / (1 - f_{\gamma})$, with $f_{\gamma} = 0.9$. The absorption factor is defined as $f_{\rm abs}(\epsilon) = 1 - \frac{1 - e^{-\tau(\epsilon)}}{\tau(\epsilon)}$, where the optical depth $\tau(\epsilon)$ is computed from Eq.~(\ref{eq:tau}). Here, $\dot{N}^0$, $\dot{N}^{\rm syn}$, and $\dot{N}^{\rm Com}$ denote the gamma-ray injection rates from primary hadronic processes, as well as from synchrotron radiation and inverse Compton scattering of secondary electrons (positrons), respectively. The steady-state electron distribution $N_{\rm e}(\gamma)$ is given as the implicit solution to Eq.~(\ref{eq:FP}):
\begin{equation}\label{eq:sol}
    N_{\rm e}(\gamma) = \frac{1}{\dot{\gamma}} \int_{\gamma}^{\infty} d\tilde{\gamma} \left\{ Q_{\rm e}(\tilde{\gamma}) + \dot{N}_{\rm e}^{\gamma\gamma}(\tilde{\gamma}) - \frac{N_{\rm e}(\tilde{\gamma})}{t_{\rm esc}} \right\}.
\end{equation}
Equation~(\ref{eq:sol}) can be solved iteratively, starting from the highest values of $\gamma$. Once the steady-state electron (positron) distribution is obtained, the resulting cascade emission can be calculated from synchrotron radiation and inverse Compton scattering.

\section{\label{sec:sph}Soft photon fields}
We can uniformly model the spectral energy distribution of soft radiation fields for all AGNs using only the SMBH mass \( M_{\bullet} \) and the observed X-ray luminosity \( L_{\rm X} \) in the 2--10 keV band~\footnote{Note that Ref.~\cite{abbasi2025icecube} selects X-ray AGNs that are particularly bright in the 20--50~keV band from the BASS catalog}, following the method proposed by Ref.~\cite{murase2020hidden}.  
The soft radiation in the corona region consists of two components: optical/UV emission from the accretion disk and X-ray emission from the corona. For the disk emission, the averaged SEDs are expressed as a function of the Eddington ratio, \( \lambda_{\rm Edd} = L_{\rm bol} / L_{\rm Edd} \) (see Figure~7 in Ref.~\cite{ho2008nuclear}), where the Eddington luminosity is $L_{\rm Edd} \approx 1.3 \times 10^{45} M_{\bullet,7}~{\rm erg~s}^{-1},$
and the bolometric luminosity can be obtained through \cite{hopkins2007observational}:
\begin{equation}\label{eq:bol}
\frac{L_{\rm bol}}{L_{\rm X}} = 10.83 \left( \frac{L_{\rm bol}}{10^{10} L_{\odot}} \right)^{0.28} + 6.08 \left( \frac{L_{\rm bol}}{10^{10} L_{\odot}} \right)^{-0.02},
\end{equation}
where $L_{\odot}$ is the solar luminosity. The disk emission is expected to cut off at a certain energy, above which the X-ray component becomes dominant.  
The cutoff energy, denoted as \( \epsilon_{\rm d} \) (also used as the characteristic disk photon energy in Sec.~\ref{sec:model}), is determined by the effective temperature at the innermost stable circular orbit (ISCO): $
T_{\rm d} \simeq 0.49 \left( \frac{G M_{\bullet} \dot{M}_{\bullet}}{72 \pi \sigma_{\rm SB} R_{\rm s}^3} \right)^{1/4},$
where \( \sigma_{\rm SB} \) is the Stefan--Boltzmann constant, and the SMBH accretion rate is  
\(\dot{M}_{\bullet} \simeq L_{\rm bol} / (\eta_{\rm rad} c^2)\) with a radiative efficiency \( \eta_{\rm rad} = 0.1 \)~\cite{kato2008black}.  The corresponding disk cutoff energy is then  $\epsilon_{\rm d} \simeq 3 k_{\rm B} T_{\rm d}.$
For the X-ray component, the SED can be modeled as a power law with an exponential cutoff:
\begin{equation}
\frac{dn_{\rm X}}{d\epsilon_{\rm X}} \propto \epsilon_{\rm X}^{-\Gamma_{\rm X}} 
\exp\!\left(-\frac{\epsilon_{\rm X}}{\epsilon_{\rm X,cut}}\right),
\end{equation}
which is normalized by the observed \( L_{\rm X} \).  
The photon index can be estimated as~\cite{trakhtenbrot2017bat}
\begin{equation}
\Gamma_{\rm X} \approx 0.167 \, \log(\lambda_{\rm Edd}) + 2,
\end{equation}
and the cutoff energy is given by~\cite{ricci2018bat}
\begin{equation}
\epsilon_{\rm X,cut} \approx -74 \, \log(\lambda_{\rm Edd}) + 150~{\rm keV}.
\end{equation}

\begin{figure}[h]
\centering
\includegraphics[width=0.6\linewidth,height=0.5\linewidth]{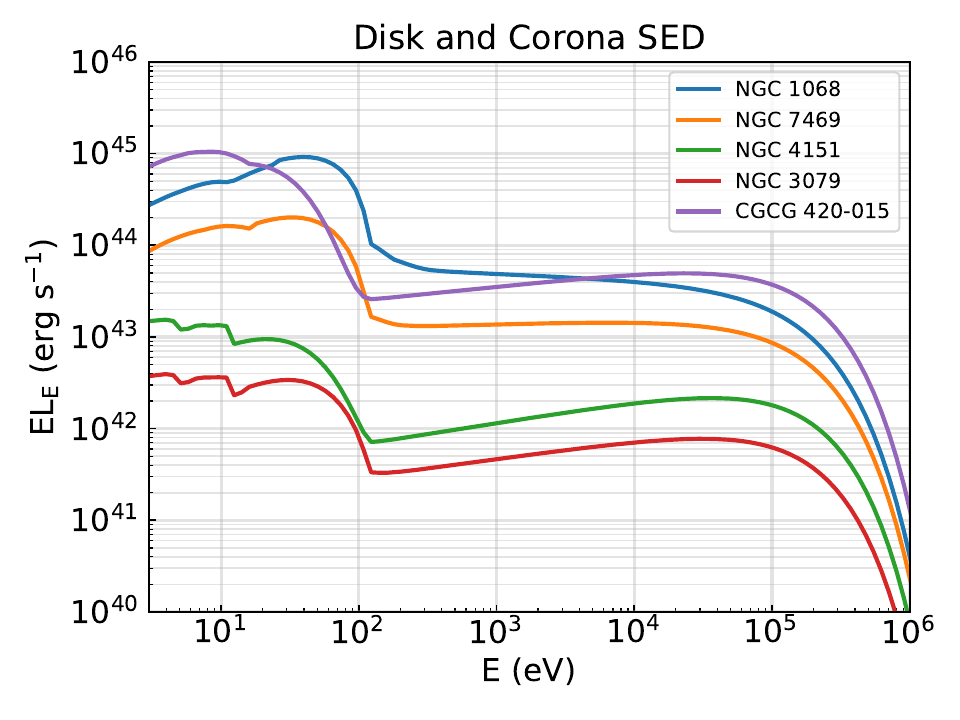}
\caption{Combined disk and corona SEDs for the five neutrino-associated Seyfert nuclei analyzed in this work.}
\label{fig:sph}
\end{figure}

Based on these relations, the combined disk--corona SED of Seyfert nuclei can be constructed once the black hole mass \( M_{\bullet} \) and X-ray luminosity \( L_{\rm X} \) are known.  
For NGC 1068, where clear observations give \( L_{\rm bol} \simeq 1 \times 10^{45}~{\rm erg~s^{-1}} \) and \( \epsilon_{\rm d} \simeq 32~{\rm eV} \), we adopt these observed values directly. For the remaining four sources, the disk--corona SEDs are derived using the relations described above. The resulting soft-photon SEDs for the five neutrino-associated Seyfert galaxies analyzed in this work are shown in Fig~\ref{fig:sph}.

\section{\label{sec:data}Data Processing}

In this work, we analyze the gamma-ray emission from NGC~1068, NGC~4151, NGC~3079, CGCG~420-015, and NGC~7469, using $\sim$ 16.4 years of Fermi-LAT observations collected between 2008 August 5 and 2025 January 1. The analysis covers the energy range from 30 MeV to 1 TeV.
Only data within a $10^{\circ}$ region of interest (ROI) centered on the position of these sources are considered. All data are retrieved from the Fermi LAT public data  \footnote{\url{https://fermi.gsfc.nasa.gov/cgi-bin/ssc/LAT/LATDataQuery.cgi}} and are processed using the Fermipy package \cite{2017ICRC...35..824W}.
We use the standard data filters: DATA\_QUAL $> 0$ and LAT\_CONFIG == 1. The photons are selected corresponding to the P8R3\_SOURCE\_V3 instrument response. The Galactic diffuse background and the point-source emission are modeled using the standard component (gll\_iem\_v07.fits) and the 4FGL-DR3 catalog (gll\_psc\_v28.fits; Ref.~\cite{abdollahi2022incremental}), respectively.
To account for photon leakage from sources outside the ROI due to the detector's point-spread function (PSF), all 4FGL sources within a $15^{\circ}$ radius are included in the model. The energy dispersion correction (edisp\_bins = -1) is applied to all sources except for the isotropic component.

Based on the energy dependence of the LAT instrument response, we divide the analysis into two energy regimes: 30-50 MeV and 50 MeV-1 TeV.
In the low energy range, the maximum zenith angle is set to $80^{\circ}$, the extragalactic emission, along with the residual instrumental background, is modeled using iso\_P8R3\_SOURCE\_V3\_v1.txt.
For the 50 MeV-1 TeV range, to optimize analysis sensitivity, we perform a joint likelihood analysis across four PSF classes (iso\_P8R3\_SOURCE\_V3\_PSFi\_v1.txt, where i ranges from 0 to 3), adopting a maximum zenith angle of $90^{\circ}$.  The data are binned using two energy bins per decade.

Before calculating the spectral energy distributions, we perform an initial model optimization. New sources with test statistics (TS) greater than 25 are identified using the Fermipy function find\_source.
The sources are modeled with a power-law spectrum. The spectral parameters (index and normalization) of both the source and the Galactic diffuse component, as well as the normalization of the isotropic component, are left free to vary. In addition, the normalization parameters of all 4FGL sources with TS $\ge 25$ located within $5^{\circ}$ of the ROI center, and of all sources with TS $\ge 500$ located within $7^{\circ}$ are free as well.
The SEDs are computed for each source using the Fermipy SED analysis, in which the flux normalization is fit independently in each energy bin, assuming a power-law spectrum with a fixed photon index of 2. Upper limits
are reported at the $95\%$ confidence level.


\nocite{*}

\acknowledgments

We acknowledge support from the National Natural Science Foundation of China under grant No.12003007 and the Fundamental Research Funds for the Central Universities (No. 2020kfyXJJS039).

\paragraph{Data Availability} The data that support the findings of this article are openly available.
\paragraph{Interest Conflict} The authors declare that they have no conflict of interest.









\bibliographystyle{unsrt}  
\bibliography{reference}        
\end{document}